\documentclass[12pt]{article}
\usepackage[dvips]{graphicx}

\usepackage{epsfig,amsmath,amssymb,verbatim,mathrsfs}
\usepackage[square, comma, sort&compress]{natbib}

\numberwithin{equation}{section}

\def\beq{\begin{eqnarray}}
\def\eeq{\end{eqnarray}}
\def\bea{\begin{eqnarray}}
\def\eea{\end{eqnarray}}

\newcommand{\gsim}{\lower.7ex\hbox{$\;\stackrel{\textstyle>}{\sim}\;$}}
\newcommand{\lsim}{\lower.7ex\hbox{$\;\stackrel{\textstyle<}{\sim}\;$}}

\def\stilde{\widetilde}

\newcommand{\newc}{\newcommand}
\newc{\Nc}{N_{c}}
\newc{\CG}{C_G}
\newc{\gp}{g'}
\newc{\stopi}{\stilde t_i}
\newc{\sboti}{\stilde b_i}
\newc{\staui}{\stilde \tau_i}
\newc{\stopj}{\stilde t_j}
\newc{\sbotj}{\stilde b_j}
\newc{\stauj}{\stilde \tau_j}
\newc{\stopI}{\stilde t_1}
\newc{\stopII}{\stilde t_2}
\newc{\sbotI}{\stilde b_1}
\newc{\sbotII}{\stilde b_2}
\newc{\stauI}{\stilde \tau_1}
\newc{\stauII}{\stilde \tau_2}
\newc{\sstop}{s_{t}}
\newc{\cstop}{c_{t}}
\newc{\ssbot}{s_{b}}
\newc{\csbot}{c_{b}}
\newc{\sstau}{s_{\tau}}
\newc{\cstau}{c_{\tau}}
\newc{\Sstop}{s_{2t}}
\newc{\Cstop}{c_{2t}}
\newc{\Ssbot}{s_{2b}}
\newc{\Csbot}{c_{2b}}
\newc{\Sstau}{s_{2\tau}}
\newc{\Cstau}{c_{2\tau}}
\newc{\salpha}{s_\alpha}
\newc{\calpha}{c_\alpha}
\newc{\Calpha}{c_{2\alpha}}
\newc{\Salpha}{s_{2\alpha}}
\newc{\sbetapm}{s_{\beta_\pm}}
\newc{\cbetapm}{c_{\beta_\pm}}
\newc{\Sbetapm}{s_{2 \beta_\pm}}
\newc{\Cbetapm}{c_{2 \beta_\pm}}
\newc{\sbetaO}{s_{\beta_0}}
\newc{\cbetaO}{c_{\beta_0}}
\newc{\SbetaO}{s_{2 \beta_0}}
\newc{\CbetaO}{c_{2 \beta_0}}
\newc{\vu}{v_u}
\newc{\vd}{v_d}
\newc{\seL}{\stilde e_L}
\newc{\smuL}{\stilde \mu_L}
\newc{\seR}{\stilde e_R}
\newc{\smuR}{\stilde \mu_R}
\newc{\suL}{\stilde u_L}
\newc{\sdL}{\stilde d_L}
\newc{\suR}{\stilde u_R}
\newc{\sdR}{\stilde d_R}
\newc{\scL}{\stilde c_L}
\newc{\ssL}{\stilde s_L}
\newc{\scR}{\stilde c_R}
\newc{\ssR}{\stilde s_R}
\newc{\snue}{\stilde \nu_e}
\newc{\snumu}{\stilde \nu_\mu}
\newc{\snutau}{\stilde \nu_\tau}
\newc{\Gpm}{G^\pm}
\newc{\Hpm}{H^\pm}
\newc{\FFbS}{\overline{FF}S}
\newc{\FFbV}{\overline{FF}V}
\newc{\FSS}{F_{SS}}
\newc{\FSSS}{F_{SSS}}
\newc{\FFFS}{F_{FFS}}
\newc{\FFFbS}{F_{\overline{FF}S}}
\newc{\FSSV}{F_{SSV}}
\newc{\FVS}{F_{VS}}
\newc{\FVVS}{F_{VVS}}
\newc{\FFFV}{F_{FFV}}
\newc{\FFFbV}{F_{\overline{FF}V}}
\newc{\Fgauge}{F_{\rm gauge}}
\newc{\DRbarprime}{$\overline{\rm DR}'$ }
\newc{\DRbar}{$\overline{\rm DR}$ }
\newc{\MSbar}{$\overline{\rm MS}$ }
\newc{\Yu}{{\bf Y}_u}
\newc{\Yd}{{\bf Y}_d}
\newc{\Ye}{{\bf Y}_e}
\newc{\Au}{{\bf a}_u}
\newc{\Ad}{{\bf a}_d}
\newc{\Ae}{{\bf a}_e}
\newc{\bm}{{\bf m}}
\newc{\zhol}{Z^{\rm hol}}

\newcommand{\Eq}[1]{Eq.\;(\ref{#1})}

\newcommand{\drawsquare}[2]{\hbox{%
\rule{#2pt}{#1pt}\hskip-#2pt
\rule{#1pt}{#2pt}\hskip-#1pt
\rule[#1pt]{#1pt}{#2pt}}\rule[#1pt]{#2pt}{#2pt}\hskip-#2pt
\rule{#2pt}{#1pt}}

\newcommand{\Yfund}{\raisebox{-.5pt}{\drawsquare{6.5}{0.4}}}
\newcommand{\Ysymm}{\raisebox{-.5pt}{\drawsquare{6.5}{0.4}}\hskip-0.4pt%
        \raisebox{-.5pt}{\drawsquare{6.5}{0.4}}}
\newcommand{\Yasymm}{\raisebox{-3.5pt}{\drawsquare{6.5}{0.4}}\hskip-6.9pt%
        \raisebox{3pt}{\drawsquare{6.5}{0.4}}}

\begin{document}

\setlength{\baselineskip}{0.2in}



\begin{titlepage}
\noindent
\begin{flushright}
\end{flushright}
\vspace{1cm}

\begin{center}
  \begin{Large}
    \begin{bf}
The Phase Structure of Supersymmetric $Sp(2 N_c)$ Gauge Theories with an Adjoint\\
     \end{bf}
  \end{Large}
\end{center}
\vspace{0.2cm}

\begin{center}

\begin{large}
David Poland\\
\end{large}
\vspace{0.3cm}
  \begin{it}
Jefferson Physical Laboratory, Harvard University,\\
Cambridge, Massachusetts 02138, USA\\
\vspace{0.5cm}
\end{it}

\end{center}

\center{\today}

\begin{abstract}
We study the phase structure of $\mathcal{N} = 1$ supersymmetric
$Sp(2N_c)$ gauge theories with $2 N_f$ fundamentals, an adjoint,
and vanishing superpotential.  Using a-maximization, we derive
analytic expressions for the values of $N_f$ below which the first
several gauge-invariant operators in the chiral ring violate the
unitarity bound and become free fields.  In doing so we are able
to explicitly check previous conjectures about the behavior of
this theory made by Luty, Schmaltz, and Terning.  We then compare
this to an analysis of the first two 'deconfined' dual
descriptions based on the gauge groups $Sp(2 N_f+2) \times
SO(2N_c+5)$ and $Sp(2N_f+2) \times SO(4N_f+4) \times Sp(2N_c+2)$,
finding precise agreement. In particular, we find no evidence for
non-obvious accidental symmetries or the appearance of a mixed
phase in which one of the dual gauge groups becomes free.
\end{abstract}

\vspace{1cm}

\end{titlepage}

\setcounter{page}{2}


\vfill\eject



\newpage

\section{Introduction}
\label{sec:intro}

It is of great interest to understand the low-energy behavior of
asymptotically free gauge theories.  Analyzing such theories in
the strong-coupling regime is generally quite difficult, but can
become tractable with a sufficient amount of symmetry.
Strongly-coupled gauge theories with $\mathcal{N} = 1$
supersymmetry are particularly interesting in that they are both
amenable to analysis and potentially relevant for phenomenology.
Possible applications include dynamical supersymmetry
breaking~\cite{Witten:1981nf,Affleck:1983rr,Poppitz:1998vd},
conformal sequestering~\cite{Luty:2001jh,Dine:2004dv}, dynamical
solutions to the
$\mu/B\mu$-problem~\cite{Roy:2007nz,Murayama:2007ge,Perez:2008ng},
dynamical explanations of the flavor
hierarchies~\cite{Nelson:2000sn,Kobayashi:2001is,Poland:2009},
dynamical solutions to the doublet-triplet splitting
problem~\cite{Yanagida:1994vq,Hotta:1996qb,Cheng:1997fk}, and so
on.

When analyzing a supersymmetric gauge theory, a first-order
question is to determine what kind of phase the theory flows to at
low energies.  Possibilities include that the theory is infrared
(IR) free, the theory flows to an interacting conformal
(non-Abelian Coulomb) phase, the theory has a 'dual' description
that is IR free, the theory confines, the theory dynamically
generates a superpotential that breaks the gauge group, or that
the theory enters a pure Abelian Coulomb phase
(see~\cite{Intriligator:1995au,Terning:2003th} for reviews).  In
addition, it is possible that the theory enters a 'mixed phase'
consisting of decoupled sectors that are in some combination of
the above phases.  This can occur, e.g., when the theory is in an
interacting conformal regime but some set of operators have become
free fields and decoupled from the CFT.  More exotically, the
theory could have a dual description containing a product gauge
group in which one gauge group is interacting and one gauge group
is IR free.  This was argued to occur, e.g., in $SU(N_c)$ gauge
theories with an anti-symmetric
tensor~\cite{Terning:1997jj,Csaki:2004uj}.

An important tool for studying theories in a conformal regime,
a-maximization, was introduced by Intriligator and Wecht
in~\cite{Intriligator:2003jj}, and further developed
in~\cite{Kutasov:2003iy,Kutasov:2003ux,Barnes:2004jj}.  The idea
is that the correct superconformal $U(1)_R$ symmetry can be
determined by maximizing \beq a(R_t) = \frac{3}{32}\left[3\,Tr
R_t^3 - Tr R_t\right] \eeq over all possible trial R-symmetries
$R_t = R_0 + \sum_I s_I F_I$, where $R_0$ is any initial
R-symmetry and $F_I$ are the IR flavor symmetries.  This is an
extremely powerful technique provided that one understands the IR
flavor symmetries.  Unfortunately, one cannot always identify the
IR flavor symmetries as a subset of the UV flavor symmetries --
accidental symmetries can arise (see, e.g.,~\cite{Leigh:1996ds} for a
number of interesting examples).

How can one gain evidence for accidental symmetries?  One way is
to check whether there are any gauge-invariant operators in the
chiral ring of the theory that violate the unitarity bound, given
by $R_{\mathcal{O}} \geq 2/3$ for scalar
operators~\cite{Mack:1975je}.  If this bound appears to be
violated, then one plausible interpretation is that $\mathcal{O}$
is becoming a free field with $R_{\mathcal{O}} = 2/3$.  In this
case there is an accidental symmetry associated with rotations of
$\mathcal{O}$, and one must include this symmetry when maximizing
$a(R_t)$.  In practice, this requires one to instead
maximize~\cite{Anselmi:1997ys,Kutasov:2003iy} \beq \widetilde{a}(R_t) = a(R_t) +
\frac{dim(\mathcal{O})}{96} \left(2 - 3
R_{\mathcal{O}}\right)^2\left(5-3 R_{\mathcal{O}}\right). \eeq

However, not all accidental symmetries manifest themselves through
apparent violations of unitarity.  This happens for example in
$SU(N_c)$ gauge theories with $N_f$ flavors of vector-like quarks
$\{\bar{Q},Q\}$ in the range $N_c + 1 < N_f < 3/2
N_c$~\cite{Seiberg:1994pq}.  Since the anomaly-free $U(1)_R$
symmetry is $R_{\bar{Q},Q} = 1 - N_c/N_f$, the mesons $\bar{Q} Q$
appear to violate the unitarity bound in this range and presumably
become free fields.  However, this is only part of the story, as
it is also believed that the entire dual $SU(N_f - N_c)$ gauge
group and the corresponding dual quarks are also becoming free
fields, yielding many more accidental symmetries.  This is not
obvious in the original 'electric' description of the theory, but
becomes apparent when the dual 'magnetic' description is analyzed.
When studying similar theories, it is clearly of great interest to
have dual descriptions available that can be studied, as they may
contain evidence for the emergence of non-obvious accidental IR
symmetries.

In the present work we will use a-maximization to study
$\mathcal{N}=1$ supersymmetric $Sp(2 N_c)$ gauge theory with $2
N_f$ fundamentals $Q_i$ and an adjoint $A$.  While the theory with
superpotential $W = A^{2(k+1)}$ is fairly well
understood~\cite{Leigh:1995qp}, the theory with vanishing
superpotential has not been as easy to analyze.  An attempt to
study this theory using 'deconfinement'~\cite{Berkooz:1995km} was
made in~\cite{Luty:1996cg}, where it was proposed that the theory
has a sequence of dual descriptions, the first of which is based
on an $Sp(2 N_f+2) \times SO(2N_c+5)$ gauge theory.  However,
because the $U(1)_R$ symmetry of the theory was unknown it was not
possible to determine which operators, if any, gave apparent
violations of the unitarity bound as one varies $N_f$ and $N_c$.
Furthermore, it was not possible to explicitly check the dual
descriptions for evidence of additional accidental symmetries, as
would occur if either of the dual gauge groups were becoming free.
As we will see, such an analysis is now possible using
a-maximization, and we will here present an attempt to map out the
phase structure of the theory.  Similar studies of other theories have appeared in~\cite{Kutasov:2003iy,Intriligator:2003mi,Csaki:2004uj,Barnes:2005zn,Kawano:2005nc,Kawano:2007rz}.

This paper is organized as follows.  In section~\ref{sec:sp2n} we
use a-maximization to study $Sp(2 N_c)$ gauge theory with an
adjoint.  In sections~\ref{sec:dual} and~\ref{sec:seconddual} we
perform a similar analysis of the first two dual descriptions of
the theory, comparing the results and looking for evidence of
accidental IR symmetries.  We give concluding remarks in
section~\ref{sec:concl}.

\section{$Sp(2 N_c)$ Gauge Theory with an Adjoint}
\label{sec:sp2n}

We are interested in studying $\mathcal{N}=1$ supersymmetric $Sp(2
N_c)$ gauge theory\footnote{Our conventions are such that $Sp(2)
\sim SU(2)$.} with $2 N_f$ fundamentals $Q_i$ and an adjoint $A$.
The field content and anomaly-free symmetries are given in
Table~\ref{fields}.
\begin{table}
\begin{center}
\begin{tabular}{| c | c | c | c | c | }
\hline
            & $SU(2N_c)$ & $SU(2N_f)$ & $U(1)_X$ & $U(1)'_R$ \\
\hline
$Q_i$        &    \Yfund     &    \Yfund        & $\frac{N_c +1}{N_f}$   & 1  \\
\hline
$A$          &    \Ysymm           &        {\bf 1}           & -1      & 0  \\
\hline
\end{tabular}
\end{center}
\caption{Field content of the theory.} \label{fields}
\end{table}
In particular, we are here interested in the theory with vanishing
superpotential.  It is believed that this theory is in an
interacting conformal regime for all $0 < N_f < 2 (N_c+1)$.

The chiral ring of this theory contains the gauge-invariant
operators \bea
T_k &\equiv& Tr A^{2 k},\,\,\, k = 1,2,... \nonumber\\
M_k &\equiv& Q A^k Q,\,\,\, k = 0,1,... \eea As mentioned in the
introduction, this theory was previously studied using
'deconfinement' in~\cite{Luty:1996cg}, where it was conjectured
that the operators $M_k$ sequentially become free fields as $N_f$
is decreased from the asymptotic freedom limit of $N_f = 2(N_c +
1)$ while the $T_k$ remain interacting.  It was also noted
in~\cite{Barnes:2005zn} that the large $N_c,N_f >> 1$ limit of
this theory will yield the same R-charges as $SU(N_c)$ gauge
theory with $N_f$ flavors and an adjoint, which was studied
in~\cite{Kutasov:2003iy}.  Here we will attempt to map out the
phase space allowing for smaller $N_f$ and $N_c$, comparing our
results to the conjectures of~\cite{Luty:1996cg}.  In particular,
we will find that both the operators $T_i$ and $M_i$ sequentially
become free fields as $N_f$ is decreased, with the precise order
depending on the value of $N_c$.  This realizes the behavior that
was described as scenario C in~\cite{Luty:1996cg}, and thus we
will prove that the conjectured behavior (scenario A) is
incorrect.

In order to determine the $U(1)_R$ symmetry of the theory, we
should maximize $a(R_{t})$ subject to the constraint that the
mixed $Tr[U(1)_R Sp(2N_c)^2]$ anomalies vanish.  Recall that an
adjoint of $Sp(2 N_c)$ is a two-index symmetric tensor with index
$(2 N_c + 2)$ and dimension $N_c (2 N_c + 1)$.  Anomaly
cancellation then requires \beq 0 = (2 N_c + 2) + 2 N_f (R_Q - 1)
+ (2 N_c + 2) (R_A - 1), \eeq or equivalently \beq R_Q = 1 -
\left(\frac{N_c+1}{N_f}\right) R_A. \eeq In order to determine the
R-symmetry, we should then maximize \bea\label{afunction}
a(R_A) &=& \frac{3}{32}\left[2N_c(2 N_c+1) +  4 N_f N_c\left(3(-\frac{N_c+1}{N_f} R_A)^3-(-\frac{N_c+1}{N_f} R_A)\right)\right. \nonumber\\
  && \qquad \left. \phantom{\frac{A}{B}} +  N_c (2 N_c +1)\left(3(R_A-1)^3 - (R_A -1)\right)\right].
\eea The correct solution to $d a / d R_A = 0$ is then given by
\beq\label{a1solution} R_A = \frac{-3(1+2N_c)N_f^2 +
\sqrt{16(1+N_c)^3(3+5N_c)N_f^2-(3+4N_c(2+N_c))N_f^4}}{12(1+N_c)^3-3(1+2N_c)N_f^2},
\eeq where the positive root of the quadratic equation is picked
out by requiring that this be a maximum.

Now we can ask the question of which gauge-invariant operator is
the first to violate the unitarity bound as we lower $N_f$ from
$2(N_c+1)$.  It is straightforward to solve the condition $R_{T_1}
\leq 2/3$ for $N_f$.  This gives the condition that the operator
$T_1$ is at or below the unitarity bound when \beq N_f \leq
2(N_c+1)\sqrt{\frac{1+N_c}{7+10N_c}}. \eeq On the other hand, we
should check that $M_0$ does not hit the unitarity bound first.
The condition $R_{M_0} \leq 2/3$ is equivalent to
 \beq N_f \leq
2(N_c+1)\left(\frac{3+6N_c-\sqrt{13+40N_c+28N_c^2}}{4+8
N_c}\right), \eeq which is less than
$2(N_c+1)\sqrt{\frac{1+N_c}{7+10N_c}}$ for positive $N_c$.  Thus,
$R_{M_0} > 2/3$ within the entire region \beq
2(N_c+1)\sqrt{\frac{1+N_c}{7+10N_c}} < N_f < 2(N_c+1). \eeq

When $N_f$ is below this threshold we assume that $T_1$ becomes a
free field, and we should modify the a-maximization procedure
according to the prescription given in~\cite{Kutasov:2003iy}.
Thus, we should now maximize the function \beq a_2(R_A) = a(R_A) +
\frac{1}{96}(2 - 6 R_A)^2(5 - 6 R_A). \eeq Again solving $d a_2 /
d R_A = 0$ yields the the solution \bea\label{a2solution}
R_A &=& \left(12N_c(1+N_c)^3-3(-8+N_c+2 N_c^2)N_f^2\right)^{-1}\left[-3(-4+N_c+2N_c^2)N_f^2 \phantom{\frac{A}{B}}\right. \\
&& \left.\phantom{\frac{A}{B}}\!\!\!\!\!\!\!\!\!\!\!\!\!\!\!\! +
\sqrt{16(1+N_c)^3(-4+N_c(3+5N_c))N_f^2+(16+N_c(40+N_c(45-4N_c(2+N_c))))N_f^4}
\right] \nonumber, \eea where the positive root is again picked
out by requiring that this be a maximum.

Now that we know the R-symmetry in this region, we can determine
which operator is next to hit the unitarity bound.  It is
straightforward to show that the condition that $R_{T_2}\leq 2/3$
is equivalent to \beq\label{t2eq} N_f \leq
2(N_c+1)\sqrt{\frac{N_c(1+N_c)}{-24+N_c(37+58 N_c)}}. \eeq This
should be compared to the condition for $R_{M_0} \leq 2/3$, which
occurs when \bea N_f \leq 2(N_c+1) \left(\frac{-12 + 3 N_c +6
N_c^2 - \sqrt{16+N_c(-88+N_c(-67+4 N_c(10+7 N_c)))}}{4 (-8+N_c+2
N_c^2)}\right).\phantom{AA} \eea It is then easily verified that
$R_{M_0} > 2/3$ in the entire region \beq
2(N_c+1)\sqrt{\frac{N_c(1+N_c)}{-24+N_c(37+58 N_c)}} < N_f \leq
2(N_c+1)\sqrt{\frac{1+N_c}{7+10N_c}}. \eeq

Below this threshold we can repeat the procedure, treating $T_2$
as a free field, and maximize \beq a_3(R_A) = a_2(R_A) +
\frac{1}{96}(2 - 12 R_A)^2(5 - 12 R_A). \eeq Solving $d a_3 / d
R_A = 0$ then yields the maximum \bea\label{a3solution}
R_A &=& \left(12 N_c(1+N_c)^3-3(-72+N_c+2 N_c^2)N_f^2\right)^{-1}\left[-3(-20+N_c+2N_c^2)N_f^2 \phantom{\frac{A}{B}}\right. \\
 &&\!\!\!\!\!\!\!\!\!\!\!\!\!\!\!\!\!\left.\phantom{\frac{A}{B}} +
\sqrt{16(1+N_c)^3(-12+N_c(3+5N_c))N_f^2+(144+N_c(522+N_c(813-4
N_c(2+N_c))))N_f^4} \right] \nonumber. \eea

Now we find that $R_{T_3} \leq 2/3$ when \bea\label{t3eq} N_f \leq
2(N_c+1)\sqrt{\frac{N_c(1+N_c)}{-144+N_c(91+146 N_c)}}, \eea and
$R_{M_0} \leq 2/3$ when \bea\label{m0decouple1} N_f \leq 2(N_c+1)
\left(\frac{-60 + 3 N_c +6 N_c^2 - \sqrt{144+N_c(-600+N_c(-323+4
N_c(10+7 N_c)))}}{4 (-72+N_c+2 N_c^2)}\right).\phantom{AA} \eea
The structure this time is somewhat more complicated.  If $N_c
\leq 28$ then $T_3$ gives the stronger bound, and if $N_c > 28$
then $M_0$ gives the stronger bound.  However, over a wide range
of $N_c$ the difference between these functions is $\lesssim 1$,
and $M_0$ and $T_3$ hit the unitarity bound at approximately the
same values of $N_f$.  In any case, since we are here primarily
interested in mapping out the phase structure for smaller values
of $N_c$ and $N_f$, we will first consider the case that $T_3$
decouples at the larger value of $N_f$.

Treating $T_3$ as a free field below the threshold in \Eq{t3eq},
we should now maximize \beq a_4(R_A) = a_3(R_A) + \frac{1}{96}(2 -
18 R_A)^2(5 - 18 R_A), \eeq which gives \bea\label{a4solution}
R_A &=& \left(12N_c(1+N_c)^3-3(-288+N_c+2 N_c^2)N_f^2\right)^{-1}\left[-3(-56+N_c+2N_c^2)N_f^2 \phantom{\frac{A}{B}}\right. \\
 &&\!\!\!\!\!\!\!\!\!\!\!\!\!\!\!\left.\phantom{\frac{A}{B}} +
\sqrt{16(1+N_c)^3(-24+N_c(3+5N_c))N_f^2+(576+N_c(2544+N_c(3933-4
N_c(2+N_c))))N_f^4} \right]. \nonumber \eea We can then determine
that $R_{T_4}\leq2/3$ when \bea N_f \leq
2(N_c+1)\sqrt{\frac{N_c(1+N_c)}{-480+N_c(169+274 N_c)}} \eea and
$R_{M_0} \leq 2/3$ when \bea\label{m0decouple2} N_f \leq 2(N_c+1)
\left(\frac{-168 +3N_c+6N_c^2 - \sqrt{576+N_c(-2064+N_c(-659+4
N_c(10+7 N_c)))}}{4 (-288+N_c+2 N_c^2)}\right).\phantom{AA} \eea
Comparing these functions, we find that $M_0$ decouples at the
larger value of $N_f$ for all $8 < N_c \leq 28$.  On the other
hand, if $N_c \leq 8$, then the bound is only potentially
applicable for $N_f = 1$. Thus, $M_0$ is the next operator to
decouple except in the special case of $N_f = 1$, and this
decoupling occurs when $N_f$ is below the threshold given in
\Eq{m0decouple2}.

As an immediate check on the results obtained so far, we can
expand Eqs.\;(\ref{m0decouple1}) and (\ref{m0decouple2}) in the
limit of large $N_c$.  Since the effect of decoupling $T_i$ can be
neglected in this limit, they should agree up to terms of
$O(1/N_c)$.  In addition, as was noted in~\cite{Barnes:2005zn},
the resulting bound on $N_f$ below which $M_0$ becomes free should
reproduce the results of~\cite{Kutasov:2003iy}.  We find these
checks to be successful.  In the large $N_c$ limit we obtain that
$M_0$ becomes a free field when \beq N_f \leq
\left(\frac{3-\sqrt{7}}{2}\right) N_c + \left(\frac32 -
\frac{17}{4\sqrt{7}}\right) + O\left(1/N_c\right),\eeq which does
indeed agree with~\cite{Kutasov:2003iy} at leading order.

\begin{figure}\label{spnplot}
\begin{center}
\includegraphics[scale=.42]{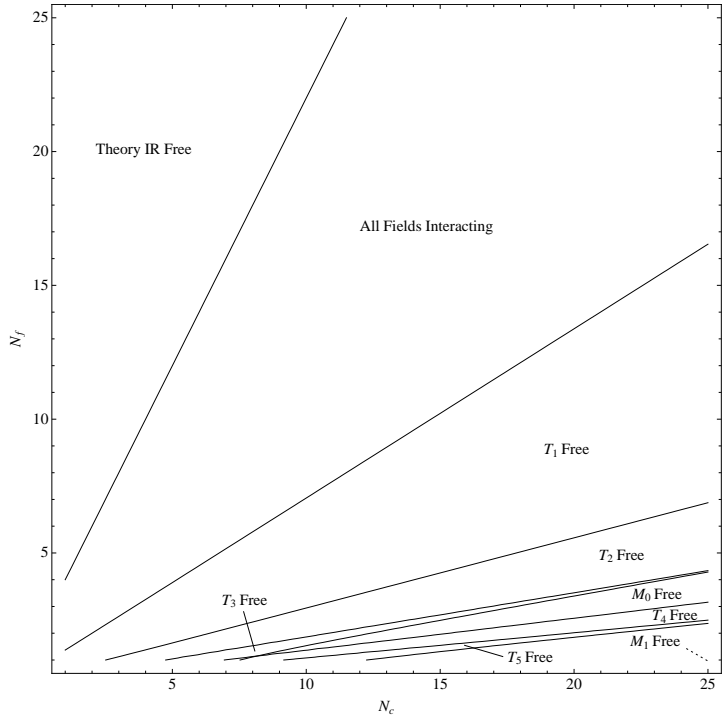}
\end{center}
\caption{Phase space diagram for $Sp(2N_c)$ with $2 N_f$
fundamentals and an adjoint with vanishing superpotential.}
\end{figure}

Of course, one can continue this procedure indefinitely.  On the
other hand, if one is primarily interested in smaller values of
$N_c$ and $N_f$, the region of interest is quickly filled in.  In
Figure~\ref{spnplot} we show an approximate phase space diagram
for the theory including the next several decoupling thresholds
(though we will suppress the analytic expressions).  Since the
case of $N_f=1$ is somewhat more complicated, we give its
structure separately in Table~\ref{nf1table}.

\begin{table}
\begin{center}
\begin{tabular}{|c|c|c|c|c|c|c|c|c|c|c|c|c|c|c|c|c|c|c|c| }

\hline
$N_c$&1&2&3&4&5&6&7&8&9&10&11&12&13&14&15&16&17&18&19                       \\
\hline
$\mathcal{O}$&&$T_1$&$T_2$&&$T_3$&&$T_4$&$M_0$&&$T_5$&&$T_6$&$M_1$&$T_7$&&$T_8$&&$M_2$&$T_9$\\
\hline
\end{tabular}
\end{center}
\caption{The values of $N_c$ in the $N_f = 1$ case for which the
operators $T_i$ and $M_i$ first become free fields.}
\label{nf1table}
\end{table}

\section{First Deconfined Dual Description}
\label{sec:dual}

One potential danger with the analysis presented in the previous
section is that it may be overlooking non-obvious accidental
symmetries that may emerge due to the strong dynamics.  This
happens, e.g., in $SU(N_c)$ gauge theories with $N_f$ flavors in
the range $N_c + 1 < N_f < 3/2 N_c$.  In that situation, there are
accidental symmetries associated with the dual quarks and
$SU(N_f-N_c)$ gauge bosons becoming free fields in the IR.  This
is manifest when the theory is studied in the dual magnetic
description, but completely non-obvious in the electric
description.  Thus, it is great interest to study dual
descriptions of the present scenario in order to look for evidence
for a similar scenario.

Fortunately, this theory in fact has a sequence of dual
descriptions that can be studied~\cite{Luty:1996cg}.  The
dualities are obtained via 'deconfinement', and involve promoting
the adjoint $A$ to a composite state in a strongly-coupled
$SO(N_c')$ theory.  The first dual description is then obtained by
taking $N_c' = 2N_c+5$ and dualizing the original $Sp(2N_c)$ gauge
group using the known duality for $Sp(2 N_c)$ gauge groups with
only fundamentals~\cite{Intriligator:1995ne}.  The end result of
this procedure is given in Table~\ref{fieldsdual}. In addition,
the theory has a superpotential \beq\label{dualW} W = M_0
\widetilde{Q} \widetilde{Q} + A_1 \widetilde{x}_1 \widetilde{x}_1
+ m_1 \widetilde{Q} \widetilde{x}_1 + m_2 \widetilde{Q}
\widetilde{p}_2 + (\widetilde{x}_1 \widetilde{p}_2)
(\widetilde{x}_1 \widetilde{p}_2) p_3, \eeq and the mapping
between the gauge-invariant operators in the original and dual
description is given by \bea\label{mapping}
Tr A^{2k} &\rightarrow& Tr A_1^{2 k} \nonumber\\
QQ &\rightarrow& M_0 \nonumber\\
Q A^k Q &\rightarrow& m_1 A_1^{k -1} m_1,\,\,\, k \geq 1. \eea

\begin{table}
\begin{center}
\begin{tabular}{| c | c | c | c | c | c |}
\hline
            & $Sp(2N_f+2)$ & $SO(2N_c+5)$ & $SU(2N_f)$ & $U(1)_X$ & $U(1)'_R$ \\
\hline
$\widetilde{Q}$        &    \Yfund     &    {\bf 1}        & $\overline{\Yfund}$  & $-\frac{N_c+1}{N_f}$  & 0\\
\hline
$M_0$              &    {\bf 1}    &    {\bf 1}         & \Yasymm              & $2 \frac{N_c+1}{N_f}$ & 2  \\
\hline
$\widetilde{x}_1$      &    \Yfund     &    \Yfund         & {\bf 1}              & $\frac12$  & 1\\
\hline
$A_1$              &    {\bf 1}    &    \Yasymm        & {\bf 1}              & $-1$            & 0  \\
\hline
$m_1$              &    {\bf 1}    &    \Yfund         & \Yfund               & $\frac{N_c+1}{N_f} - \frac12$  & 1\\
\hline
$m_2$              &    {\bf 1}    &    {\bf 1}       & \Yfund                & $\frac{N_c+1}{N_f} + \frac12 - N_c$  & 5  \\
\hline
$\widetilde{p}_2$      &    \Yfund     &    {\bf 1}        & {\bf 1}              & $-\frac12 + N_c$  & -3\\
\hline
$p_3$              &    {\bf 1}    &    {\bf 1}        &  {\bf 1}             & $- 2 N_c$      & 6  \\
\hline
\end{tabular}
\end{center}
\caption{Field content of the dual theory.} \label{fieldsdual}
\end{table}

Now, if one just considers the one-loop beta function it na\"\i
vely seems that the $SO(2N_c+5)$ gauge group is IR free for $N_f
\geq N_c+1$.  However, this is misleading because, e.g., the
strong $Sp(2N_f+2)$ gauge group gives a large anomalous dimension
to the bi-fundamental $\widetilde{x}_1$, which in turn gives an
$O(1)$ correction to the full $SO(2N_c+5)$ beta function.  This is
an example of a larger class of RG flows in product group theories
(see~\cite{Poppitz:1996vh,Barnes:2005zn} for many examples) in
which an otherwise IR free coupling can be driven to be
interacting because of the other gauge group.

Thus, we will proceed by assuming that the full theory is
interacting, again using a-maximization to decide if any fields
become free.  In particular, we should require that the $U(1)_R$
symmetry is anomaly free with respect to both gauge groups, i.e.
that both the $Tr[U(1)_R Sp(2N_f+2)^2]$ and $Tr[U(1)_R
SO(2N_c+5)^2]$ anomalies vanish.  Furthermore, we will start by
assuming that the full superpotential in \Eq{dualW} is marginal.
Notice that anomaly cancellation and the superpotential together
give 7 constraints on 8 unknown $U(1)_R$ charges, causing
$a^{dual}(R_i)$ to again be a function of a single variable as in
the previous section.

More concretely, the constraints are
\bea
0 &=& (2N_f+4) + 2 N_f(R_{\widetilde{Q}}-1) + (2N_c+5)(R_{\widetilde{x}_1} - 1) + (R_{\widetilde{p}_2} - 1) \nonumber\\
0 &=& (2N_c+3) + (2N_f+2)(R_{\widetilde{x}_1} - 1) + (2N_f)(R_{m_1}-1) + (2N_c+3)(R_{A_1}-1) \nonumber\\
2 &=& R_{M_0}+2R_{\widetilde{Q}} \nonumber\\
2 &=& R_{A_1}+2R_{\widetilde{x}_1} \nonumber\\
2 &=& R_{m_1}+R_{\widetilde{Q}}+R_{\widetilde{x}_1} \nonumber\\
2 &=& R_{m_2}+R_{\widetilde{Q}}+R_{\widetilde{p}_2} \nonumber\\
2 &=& 2 R_{\widetilde{x}_1} + 2 R_{\widetilde{p}_2}+ R_{p_3}. \eea
These linear equations can be solved to express all of the
R-charges in terms of $R_{A_1}$, yielding \bea\label{rcharges}
R_{\widetilde{Q}} &=& \frac{1+N_c}{N_f} R_{A_1} \nonumber\\
R_{M_0}      &=& 2 - 2\frac{1+N_c}{N_f} R_{A_1} \nonumber\\
R_{\widetilde{x}_1} &=& 1 - \frac12 R_{A_1} \nonumber\\
R_{m_1}      &=& 1 + \frac{N_f - 2(1+N_c)}{2 N_f} R_{A_1} \nonumber\\
R_{m_2}      &=& 5 - \frac{N_f + 2 - 2N_c(N_f -1)}{2N_f} R_{A_1} \nonumber\\
R_{\widetilde{p}_2} &=& -3 + \frac{1-2N_c}{2} R_{A_1} \nonumber\\
R_{p_3}        &=& 6 + 2 N_c R_{A_1}. \eea Equivalently, we could
have obtained these expressions simply by considering the linear
combination $R_i = R'_i - R_{A_1} X_i$, where $X_i$ and $R'_i$ are
the $U(1)_X$ and $U(1)_R'$ charges of each field as given in
Table~\ref{fieldsdual}.

In order to determine $R_{A_1}$ we should then maximize the
function \bea\label{dualfunction}
a^{dual}(R_{A_1}) &=& \frac{3}{32}\left[2(N_f+1)(2 N_f+3)+2(N_c+2)(2 N_c+ 5)  \phantom{\frac12}\right. \nonumber\\
  && \qquad     +  (N_c+2)(2N_c+5)\left(3(R_{A_1}-1)^3 - (R_{A_1} -1)\right) \nonumber\\
  && \qquad     +  (2N_f+2)(2N_f)\left(3(\frac{1+N_c}{N_f} R_{A_1}-1)^3-(\frac{1+N_c}{N_f} R_{A_1}-1)\right)            \nonumber\\
  && \qquad     +  (N_f)(2 N_f - 1)\left(3(1 - 2\frac{1+N_c}{N_f} R_{A_1})^3 - (1 - 2\frac{1+N_c}{N_f} R_{A_1})\right)  \nonumber\\
  && \qquad     +  (2N_f+2)(2 N_c + 5)\left(3(- \frac12 R_{A_1})^3 - (- \frac12 R_{A_1})\right) \nonumber\\
  && \qquad     +  (2N_c+5)(2 N_f)\left(3(\frac{N_f - 2(1+N_c)}{2 N_f} R_{A_1})^3 - (\frac{N_f - 2(1+N_c)}{2 N_f} R_{A_1})\right) \nonumber\\
  && \qquad     +  (2N_f)\left(3(4 - \frac{N_f + 2 - 2N_c(N_f -1)}{2N_f} R_{A_1})^3 - (4 - \frac{N_f + 2 - 2N_c(N_f -1)}{2N_f} R_{A_1})\right) \nonumber\\
  && \qquad     +  (2N_f+2)\left(3(-4 + \frac{1-2N_c}{2} R_{A_1})^3 - (-4 + \frac{1-2N_c}{2} R_{A_1})\right) \nonumber\\
  && \qquad \left.\!   + \left(3 (5 + 2 N_c R_{A_1})^3 - (5 + 2 N_c R_{A_1})\right)   \phantom{\frac12} \!\!\!\!\!\right].
\eea Remarkably, the solution to $da^{dual} / dR_{A_1} = 0$
corresponding to the maximum is \beq R_{A_1} =
\frac{-3(1+2N_c)N_f^2 +
\sqrt{16(1+N_c)^3(3+5N_c)N_f^2-(3+4N_c(2+N_c))N_f^4}}{12(1+N_c)^3-3(1+2N_c)N_f^2},
\eeq which coincides exactly with \Eq{a1solution}!  Moreover, the
function \Eq{dualfunction} is precisely equal to \Eq{afunction}.
Of course, one may view this as a consequence of the fact that the
't Hooft anomalies of the dual description were designed to match
those of the original theory.  Nevertheless, we can perhaps view
this agreement as a non-trivial check of the {\it dynamical}
assumptions that both gauge groups are interacting and that the
full superpotential \Eq{dualW} is marginal, at least for the
values of $N_f$ that do not lead to violations of the unitarity
bound.

Since the function $a^{dual}(R_{A_1})$ is the same as before, the
operators $Tr A_1^k$ hit the unitarity bound at the same
thresholds as before.  In particular, $Tr A_1^2$ becomes a free
field for $N_f \leq 2(N_c+1)\sqrt{\frac{1+N_c}{7+10N_c}}$.  Below
this threshold we can maximize \beq a^{dual}_2(R_{A_1}) =
a^{dual}(R_{A_1}) + \frac{1}{96}(2 - 6 R_{A_1})^2(5 - 6 R_{A_1}).
\eeq This then yields the same result as \Eq{a2solution}.  The
precise agreement between the dual description and the original
description then continues, with $Tr A_1^4$ becoming free at the
threshold given in \Eq{t2eq} and $Tr A_1^6$ becoming free at the
threshold given in \Eq{t3eq} (for $N_c \leq 28$).

Now, one might worry that the situation changes when $M_0$
violates the unitarity bound and becomes a free field.  This
occurs either below the threshold given in \Eq{m0decouple1} or
\Eq{m0decouple2}, depending on the value of $N_c$ (or at $N_f=1$
for $N_c = 8$). When this happens, the superpotential term $M_0
\widetilde{Q} \widetilde{Q}$ must be flowing to zero, since this
is $M_0$'s only interaction.  In addition, for $N_f$ just above
this threshold, we know that $R_{\widetilde{Q}} \approx 2/3$ due
to the superpotential interaction.  A possible interpretation of
this is that the $Sp(2 N_f + 2)$ gauge group is becoming free.
Under this interpretation, the coupling is flowing to zero because
unitarity is now enforcing the condition that $R_{\widetilde{Q}}>
2/3$. Note that in this case one would still expect the
$SO(2N_c+5)$ gauge group to be strongly coupled.  If this
interpretation is correct, this would be similar to the mixed
phase argued to exist in $SU(N_c)$ gauge theories with an
anti-symmetric tensor~\cite{Terning:1997jj,Csaki:2004uj}.  In
addition, there would be accidental symmetries emerging that would
invalidate the a-maximization analysis performed in the electric
description of the theory.

However, we will now argue that this scenario cannot be correct.
To do this we will consider the sign of the $Sp(2 N_f +2)$
$\beta$-function in the hypothetical mixed phase.  As mentioned
above, since we are assuming that $Sp(2 N_f + 2)$ is becoming free
we must now have $R_{\widetilde{Q}} > 2/3$ and
$R_{\widetilde{p}_2}
> 2/3$ by unitarity, and thus we know that the couplings
$m_2 \widetilde{Q} \widetilde{p}_2$ and
$(\widetilde{x}_1\widetilde{p}_2)(\widetilde{x}_1\widetilde{p}_2)p_3$
must also become irrelevant.  The interacting sector of the theory
then simply consists of the fields
$\{\widetilde{x}_1,A_1,m_1,\widetilde{Q}\}$, with superpotential
\beq\label{wmixed} W_{mixed} = A_1 \widetilde{x}_1 \widetilde{x}_1
+ m_1 \widetilde{Q} \widetilde{x}_1.\eeq

Now, in order for this scenario to be plausible the $Sp(2 N_f +2)$
$\beta$-function should be positive so that the $g_{Sp}
\rightarrow 0$ fixed point is IR attractive.  This then requires
that $Tr[U(1)_R Sp(2N_f+2)^2] < 0$, or more explicitly \beq
(2N_f+4) +(2N_f)(R_{\widetilde{Q}}-1)
+(2N_c+5)(R_{\widetilde{x}_1}-1) + (2/3-1) < 0. \eeq Because the
superpotential and anomaly cancellation constraints lead to the
same parametrization of the R-charges (for the interacting fields)
as was given in \Eq{rcharges}, this condition is equivalent to
\beq\label{ra1bound} R_{A_1} < -\frac{2}{3}
\left(\frac{11}{2N_c-1}\right).\eeq  Since we expect the theory to
have $R_{A_1} > 0$ so as to avoid an infinite number of free
operators, this bound will never be satisfied.  For example, in
the limit of large $N_c$ we obtain \beq R_{A_1} \simeq
\frac{\sqrt{5}}{3} \frac{N_f}{N_c} + O(1/N_c^2), \eeq and the full
calculation gives qualitatively similar results. Note that it can
be also verified that no subset of the couplings in
Eq.~\ref{wmixed} leads to an IR stable fixed point.  Thus, we
conclude that $g_{Sp} \rightarrow 0$ is not an IR attractive fixed
point in the hypothetical mixed phase, and that both gauge groups
must remain interacting even after $M_0$ becomes free.\footnote{It
is also interesting to note that entering the hypothetical mixed
phase would have required violating the (stronger) conjecture of
Ref.~\cite{Intriligator:2005if} that operators with $R > 5/3$
cannot become free fields, since $R_{p_3} > 6$ in the
interacting scenario.}

\section{Second Deconfined Dual Description}
\label{sec:seconddual}

Now we will consider the second dual description constructed
in~\cite{Luty:1996cg}, which can be obtained by treating the
anti-symmetric tensor $A_1$ as a meson of a confining $Sp(2N_c+2)$
gauge theory, and then dualizing the $SO(2N_c+5)$ gauge group
using the known duality for $SO(N)$ gauge theories with
fundamentals~\cite{Seiberg:1994pq,Intriligator:1995id}.  The field
content (after integrating out massive fields) is given in
Table~\ref{fieldsseconddual}.  In addition, the theory has the
superpotential \bea\label{seconddualW} W &=& M_0
(\widetilde{\widetilde{x}}_1
\widetilde{m}_1)(\widetilde{\widetilde{x}}_1 \widetilde{m}_1) +
(\widetilde{\widetilde{x}}_1
\widetilde{x}_2)(\widetilde{\widetilde{x}}_1 \widetilde{x}_2)+ m_2
\widetilde{p}_2
(\widetilde{\widetilde{x}}_1 \widetilde{m}_1) +n_1 \widetilde{p}_2^2 p_3 \nonumber\\
&& + n_1 \widetilde{\widetilde{x}}_1 \widetilde{\widetilde{x}}_1 +
A_2 \widetilde{x}_2 \widetilde{x}_2 + M_1 \widetilde{m}_1
\widetilde{m}_1 + n_3 \widetilde{r}_2
\widetilde{r}_2  \nonumber\\
&& + n_2 \widetilde{x}_2 \widetilde{m}_1 +n_4
\widetilde{\widetilde{x}}_1 \widetilde{r}_2 + n_5 \widetilde{m}_1
\widetilde{r}_2, \eea and the gauge-invariant operators of the
electric theory match onto the operators \bea\label{2ndmapping}
Tr A^{2k} &\rightarrow& Tr A_2^{2 k} \nonumber\\
QQ &\rightarrow& M_0 \nonumber\\
Q A Q &\rightarrow& M_1\nonumber\\
Q A^k Q &\rightarrow& n_2 A_2^{k -2} n_2,\,\,\, k \geq 2. \eea

\begin{table}
\begin{center}
\begin{tabular}{| c | c | c | c | c | c | c |}
\hline
            & $Sp(2N_f+2)$ & $SO(4N_f+4)$ & $Sp(2N_c+2)$ & $SU(2N_f)$ & $U(1)$ & $U(1)'_R$ \\
\hline
$M_0$              &    {\bf 1}    &    {\bf 1}         &   {\bf 1}     &    \Yasymm              & $2 \frac{N_c+1}{N_f}$ & 2  \\
\hline
$\widetilde{\widetilde{x}}_1$      &    \Yfund     &    \Yfund     &   {\bf 1}    & {\bf 1}              & $-\frac12$  & 0\\
\hline
$n_1$              &    \Ysymm    &    {\bf 1}         &  {\bf 1}        & {\bf 1}              & $1$            & 2  \\
\hline
$\widetilde{x}_2$      &    {\bf 1}    &    \Yfund         & \Yfund         & {\bf 1}      & $\frac12$      & 1\\
\hline
$A_2$              &    {\bf 1}    &    {\bf 1}       & \Ysymm         & {\bf 1}       & $-1$  & 0  \\
\hline
$\widetilde{m}_1$      &    {\bf 1}      &    \Yfund     &    {\bf 1}        & $\overline{\Yfund}$              & $\frac12 -  \frac{N_c+1}{N_f}$  & 0\\
\hline
$M_1$              &    {\bf 1}    &    {\bf 1}        &  {\bf 1}       & \Ysymm           & $2 \frac{N_c+1}{N_f} - 1$      & 2  \\
\hline
$n_2$              &    {\bf 1}    &    {\bf 1}        &  \Yfund       & \Yfund          & $\frac{N_c+1}{N_f} - 1$      & 1  \\
\hline
$n_3$              &    {\bf 1}    &    {\bf 1}        &  {\bf 1}       & {\bf 1}          & $-2 N_c - 4$      & 0  \\
\hline
$n_4$              &    \Yfund    &    {\bf 1}        &  {\bf 1}       & {\bf 1}          & $-N_c - \frac32$      & 1  \\
\hline
$n_5$              &    {\bf 1}    &    {\bf 1}        &  {\bf 1}       & \Yfund          & $\frac{N_c+1}{N_f}-\frac52-N_c$      & 1  \\
\hline
$m_2$              &    {\bf 1}    &    {\bf 1}        &  {\bf 1}       & \Yfund          & $\frac{N_c+1}{N_f}+\frac12-N_c$      & 5  \\
\hline
$\widetilde{p}_2$              &    \Yfund    &    {\bf 1}        &  {\bf 1}       & {\bf 1}          & $-\frac12+N_c$      & -3  \\
\hline
$p_3$              &    {\bf 1}    &    {\bf 1}        &  {\bf 1}       & {\bf 1}          & $-2N_c$      & 6  \\
\hline
$\widetilde{r}_2$              &    {\bf 1}    &    \Yfund        &  {\bf 1}       & {\bf 1}          & $N_c+2$      & 1  \\
\hline
\end{tabular}
\end{center}
\caption{Field content of the second dual description.}
\label{fieldsseconddual}
\end{table}

We can again begin by assuming that each of the $Sp(2N_f+2) \times
SO(4N_f+4) \times Sp(2N_c+2)$ gauge groups are interacting and
that the entire superpotential in Eq.~\ref{seconddualW} is
marginal. This gives 3 constraints from anomaly cancellation and
11 constraints from the superpotential on 15 unknown R-charges,
causing $a^{dual_2}(R_i)$ to again be a function of a single
variable (as expected).  The R-charges of each field are then
easily obtained in terms of $R_{A_2}$ by considering the linear
combination \beq\label{rchargesseconddual} R_i[R_{A_2}] = R'_i -
X_i R_{A_2},\eeq where $R'_i$ and $X_i$ are the $U(1)_R'$ and
$U(1)_X$ charges given in Table~\ref{fieldsseconddual}.
Alternatively, this parametrization could be obtained by using the
14 constraints to solve for the 15 unknown R-charges in terms
$R_{A_2}$, as we did in the previous section.

It is then straightforward to verify that the function \bea
a^{dual_2}(R_{A_2}) &=& \frac{3}{32}\left[2(N_f+1)(2 N_f+3) +
2(2N_f+2)(4 N_f+ 3) + 2(N_c+1)(2 N_c+3) \phantom{\sum_i}\right. \nonumber\\
&& \qquad \left.\!   + \sum_i \text{dim}_{\mathcal{O}_i} \left(3 (R_i[R_{A_2}] -1)^3 -
(R_i[R_{A_2}]-1)\right) \right] \eea is exactly equal to the
functions given in Eqs.~(\ref{afunction})
and~(\ref{dualfunction}).  In particular, maximizing it gives rise
to the same $U(1)_R$ symmetry as was found in the previous
sections.  Furthermore, the operators $Tr A_2^k$ become free
fields at the same thresholds as before as we lower $N_f$ from
$2(N_c+1)$.\footnote{It is perhaps worrisome that the gauge-invariant operator
$n_4 \widetilde{p}_2$ appears to badly violate the unitarity bound,
since it has $R_{n_4 \widetilde{p}_2} = -2 + 2 R_{A_2}$.  However, if the
duality is to be believed, non-perturbative effects in this description
should cause this operator to be zero in the chiral ring in order to avoid
a contradiction.}

When the operator $M_0$ hits the unitarity bound we assume that
the coupling $M_0 (\widetilde{\widetilde{x}}_1
\widetilde{m}_1)(\widetilde{\widetilde{x}}_1 \widetilde{m}_1)$ is
simply flowing to zero so that $M_0$ can become a free field. Note
that this is unlikely be the result of the $SO(4N_f+4)$ gauge
coupling flowing to zero because this would also force $M_1$ to be
a free field and one would expect that $R_{M_1} \approx 2/3$ close
to this threshold, which is not the case.  Furthermore, the
$Sp(2N_c+2)$ gauge group going free would not cause this operator
to become irrelevant.

Thus we first consider the possibility that, similar to the
hypothetical mixed phase considered in the previous section, the
$Sp(2N_f+2)$ gauge group is becoming free when $M_0$ hits the
unitarity bound.  If this is the case, unitarity requires that
$R_{n_1}, R_{n_4}, R_{\widetilde{p}_2} > 2/3$ and forces the
couplings $m_2 \widetilde{p}_2 (\widetilde{\widetilde{x}}_1
\widetilde{m}_1)$ and  $n_1 \widetilde{p}_2^2 p_3$ to become
irrelevant.  The interacting superpotential of the mixed phase
then becomes \bea\label{wmixedsecdual} W_{mixed} &=&
(\widetilde{\widetilde{x}}_1
\widetilde{x}_2)(\widetilde{\widetilde{x}}_1 \widetilde{x}_2)+
 n_1 \widetilde{\widetilde{x}}_1 \widetilde{\widetilde{x}}_1 +
A_2 \widetilde{x}_2 \widetilde{x}_2 + M_1 \widetilde{m}_1
\widetilde{m}_1 \nonumber \\
&& +\, n_3 \widetilde{r}_2 \widetilde{r}_2 + n_2 \widetilde{x}_2
\widetilde{m}_1 + n_4 \widetilde{\widetilde{x}}_1 \widetilde{r}_2
+ n_5 \widetilde{m}_1 \widetilde{r}_2, \eea along with the free
fields $\{M_0, m_2, \widetilde{p}_2, p_3\}$ and potentially free
operators $Tr A_2^{2 k}$.

The superpotential combined with anomaly cancellation then give 10
constraints on 11 unknown R-charges, with the same parametrization
for the R-charges of the interacting fields as in
Eq.~\ref{rchargesseconddual}.  However, again we can rewrite the
condition that $Tr[U(1)_R Sp(2N_f+2)^2] < 0$ as
\beq\label{ra2bound} R_{A_2} < -\frac{2}{3}
\left(\frac{11}{2N_c-1}\right),\eeq and this scenario is
disfavored for the same reason as in first dual.  It is
straightforward to additionally verify that no subset of the
couplings in Eq.~\ref{wmixedsecdual} lead to an IR attractive
fixed point.

Next we would like to investigate the possibility that when $M_1$
becomes a free field the $SO(4N_f+4)$ gauge coupling is flowing to
zero.  Note that in this case unitarity is forcing the $M_1
\widetilde{m}_1 \widetilde{m}_1$ coupling to become irrelevant. To
see if this is plausible we can again attempt to determine the
$U(1)_R$ symmetry of the hypothetical mixed phase and check the
sign of the $SO(4N_f+4)$ $\beta$-function.

If it is correct that the $SO(4N_f+4)$ gauge coupling flows to
zero, unitarity also requires that $R_{\widetilde{r}_2}>2/3$ in
addition to $R_{\widetilde{m}_1}>2/3$.  These conditions imply
that the couplings $m_2 \widetilde{p}_2
(\widetilde{\widetilde{x}}_1 \widetilde{m}_1)$, $n_3
\widetilde{r}_2 \widetilde{r}_2$, and $n_5 \widetilde{m}_1
\widetilde{r}_2$ should become irrelevant, and thus it is
reasonable to assume that the interacting superpotential becomes
\bea\label{seconddualWmixed} W_{mixed} &=&
(\widetilde{\widetilde{x}}_1
\widetilde{x}_2)(\widetilde{\widetilde{x}}_1 \widetilde{x}_2) +
n_1 \widetilde{p}_2^2 p_3 + n_1 \widetilde{\widetilde{x}}_1
\widetilde{\widetilde{x}}_1 + A_2 \widetilde{x}_2 \widetilde{x}_2
+ n_2 \widetilde{x}_2 \widetilde{m}_1 +n_4
\widetilde{\widetilde{x}}_1 \widetilde{r}_2, \eea along with the
free fields $\{M_0,M_1,n_3,n_5,m_2\}$ and potentially free
operators $Tr A_2^{2k}$.  Since we now have 8 constraints and 10
unknown R-charges, $a(R_i)$ will be a function of 2 variables and
is best maximized numerically.  Doing this, however, we find that
the function has no stable maximum and thus the fixed point does
not exist.

\begin{figure}\label{traceplot}
\begin{center}
\includegraphics[scale=1.0]{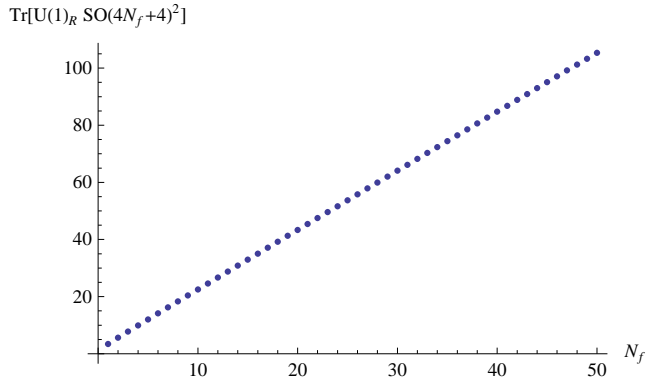}
\end{center}
\caption{$Tr[U(1)_R SO(4N_f+4)^2]$ as a function of $N_f$ in the
large $N_c$ limit of the hypothetical mixed phase.  Because it is
always positive the $SO(4N_f+4)$ gauge coupling can not flow to
zero and the mixed phase does not exist.}
\end{figure}

If we turn off one additional coupling, we find that the only IR
stable choice is to assume that
$(\widetilde{\widetilde{x}}_1\widetilde{x}_2)(\widetilde{\widetilde{x}}_1
\widetilde{x}_2)$ is flowing to zero -- i.e., only this operator
has $R > 2$ in the hypothetical CFT in which it is absent from the
superpotential.  However, in this case one can then check that
$Tr[U(1)_R SO(4N_f+4)] > 0$ for all $N_c$ and $N_f$, and hence the
assumption that $g_{SO} \rightarrow 0$ is not correct.  To
illustrate this, in Figure~\ref{traceplot} we plot $Tr[U(1)_R
SO(4N_f+4)^2]$ as a function of $N_f$ in the limit of large $N_c$.
 We have also checked that no subset of these couplings leads to an
IR stable fixed point.  We thus do not find any evidence for a
mixed phase in this description of the theory.

\section{Conclusions}
\label{sec:concl}

In this work we have attempted to map out the phase structure of
supersymmetric $Sp(2 N_c)$ gauge theories with $2 N_f$
fundamentals, an adjoint, and vanishing superpotential.  The IR
behavior of this theory has an incredibly rich structure and has
previously been difficult to analyze.  Using a-maximization,
however, we have been able to check the conjectures
of~\cite{Luty:1996cg} as well as look for evidence that the theory
enters a mixed phase below some value of $N_f$.  We have not found
any such evidence in the simplest known dual descriptions of the
theory.  It is thus tempting to believe (though far from proven)
that the original electric description of the theory is a good
description for all $N_c$ and $N_f$.

A straightforward extension of the present work would be to
construct the deconfined dual descriptions of the $SU(N_c)$
version of this theory and perform a similar analysis.  It would
also be quite interesting to find dual descriptions of these
theories in which the operators $Tr A^{2 k}$ appear as elementary
fields so that one could better understand the way in which they
decouple from the theory.  More generally, it would be interesting
to find new examples of theories that possess mixed phases (as
in~\cite{Csaki:2004uj}) so that one could better understand and
classify the situations under which they can occur.  These
possible directions are left to future work.

\section*{Acknowledgements}

We thank Jonathan Heckman, Ken Intriligator, John Mason, David Morrissey,
and David Simmons-Duffin for helpful discussions and inspiration.
This work is supported in part by the Harvard Center
for the Fundamental Laws of Nature and by NSF grant PHY-0556111.



\end{document}